\documentclass[acmsmall]{acmart}
\usepackage{booktabs}
\usepackage{tikz}
\usepackage{algorithm2e}
\usepackage{multirow}
\usepackage{csquotes}
\usepackage{tabularx}
\usepackage{graphicx}
\usepackage{caption}
\usepackage{subcaption}
\usepackage{xspace}
\usepackage{listings}
\usepackage{colortbl}
\usepackage{microtype}
\usepackage[inline,shortlabels]{enumitem}
\newcommand\qnn[2]{\ensuremath{W^{#1}A^{#2}}\xspace}
\newcommand{\finnr}{\mbox{FINN-\textsl{R}}\xspace}

\newcommand\ralg[1]{Alg.\,\ref{#1}}
\newcommand\req[1]{Eq.\,(\ref{#1})}
\newcommand\rfig[1]{Fig.\,\ref{#1}}
\newcommand\rsec[1]{Sec.\,\ref{#1}}
\newcommand\rtab[1]{Tab.\,\ref{#1}}
\makeatletter
\newif\if@internal\@internaltrue
\newsavebox\internal@box
\newlength \internal@len
               {\begin{lrbox}{\internal@box}%
                   \setlength  \internal@len{#1\linewidth}%
                   \addtolength\internal@len{-2\fboxsep}%
                   \addtolength\internal@len{-2\fboxrule}%
                   \begin{minipage}{\internal@len}}%
               {\end{minipage}\end{lrbox}%
                 \if@internal\par\noindent\fcolorbox{blue}{yellow!30}{\usebox{\internal@box}}\fi}
\newcommand\TODO[1]{\if@internal\par\noindent{\color{red}\textbf{TODO}: #1}\fi}
\makeatother

\acmJournal{TRETS}
\title[The \finnr Framework]
      {\finnr: An End-to-End Deep-Learning Framework for Fast Exploration of Quantized Neural Networks}

\author{Michaela Blott}
\email{michaela.blott@xilinx.com}
\author{Thomas B. Preu{\ss}er}
\position{Marie-Sk{\l}odowska-Curie-Fellow}
\email{thomas.preusser@utexas.edu}
\author{Nicholas J. Fraser}
\email{nfraser@xilinx.com}
\author{Giulio Gambardella}
\email{giulio.gambardella@xilinx.com}
\author{Kenneth O'Brien}
\email{kenneth.obrien@xilinx.com}
\author{Yaman Umuroglu}
\email{yamanu@xilinx.com}
\affiliation{
  \institution{Xilinx Research}
  \city{Dublin}
  \country{Ireland}
}

\author{Miriam Leeser}
\email{mel@coe.neu.edu}
\affiliation{
  \institution{Northeastern University}
  \city{Boston}
  \state{Massachusetts}
  \country{US}
}

\author{Kees Vissers}
\email{kees.vissers@xilinx.com}
\affiliation{
  \institution{Xilinx Research}
  \city{San Jos{\'e}}
  \country{US}
}

\begin{document}

\begin{abstract}
Convolutional Neural Networks have rapidly become the most successful
machine learning algorithm, enabling ubiquitous machine vision
and intelligent decisions on even embedded computing-systems.
While the underlying arithmetic is structurally simple,
compute and memory requirements are challenging.
One of the promising opportunities is leveraging reduced-precision representations
for inputs, activations and model parameters.
The resulting scalability in performance, power efficiency and storage footprint
provides interesting design compromises in exchange for a small reduction in accuracy.
FPGAs are ideal for exploiting low-precision inference engines leveraging custom
precisions to achieve the required numerical accuracy for a given application.
In this article, we describe the second generation of the FINN
framework, an end-to-end tool which enables design space exploration and
automates the creation of fully customized inference engines on FPGAs.
Given a neural network description, the tool optimizes for given platforms,
design targets and a specific precision.
We introduce formalizations of resource cost functions and performance predictions,
and elaborate on the optimization algorithms.
Finally, we evaluate a selection of reduced precision neural networks ranging from
CIFAR-10 classifiers to YOLO-based object detection on a range of platforms including PYNQ and AWS\,F1, demonstrating
new unprecedented measured throughput at 50\,TOp/s on AWS\,F1 and
5\,TOp/s on embedded devices.

\end{abstract}

\maketitle

\section{Introduction}\label{secIntroduction}
Deep Neural Networks (DNNs) achieve impressive results in computer
vision, speech recognition and many other applications.
However, the
enormous compute and storage requirements associated with their deployment
encounter limitations in regards to cost, power budget,
throughput, and latency. Energy
costs and data transmission overheads of offloading to the cloud are unsuitable
for low-latency, real-time or safety-critical application requirements, such as
speech recognition, augmented reality, drone control or autonomous
driving.
Also in the cloud
computing context itself, we face ever-increasing throughput
requirements to process astronomical scales of
data, bringing additional challenges in energy efficiency to minimize
operating expenses. While cloud service latency is less critical compared to embedded
scenarios, it still translates directly into customer experience for
interactive applications. For instance, Jouppi
et\,al. \cite{jouppi2017datacenter} quote a response time limit of
7\,ms for an interactive user experience in cloud-based services.

Neural networks using floating point operations are the initial choice from the machine learning community, but trained
parameters can contain a lot of redundant information \cite{DBLP:journals/corr/HanMD15}.
This can be exploited to reduce the computational
cost of inference. Competitive levels of accuracy have been demonstrated
using sparsification \cite{sparsecnn}, singular value decomposition
\cite{svdconvnet}, pruning \cite{DBLP:journals/corr/HanMD15,
faraone2017compressing, yu2017scalpel}, or a combination of techniques
\cite{han2015learning, iandola2016squeezenet}.
Another promising key approach for exploiting redundancy in neural networks is to use \emph{quantization}.
It has been demonstrated that moving from floating-point
arithmetic to low-precision integer arithmetic only impacts the
network accuracy lightly, especially if the network is retrained \cite{sung:2015}.
The quantization of neural networks down to 8-bit integers has
been widely adopted and is well supported by designated and optimized
software libraries, such as gemmlowp \cite{gemmlowp} and the ARM
Compute Library \cite{arm-compute}.
In this paper, we are interested in even further
reduced precisions all the way down to the extreme case of
Binary Neural Networks (BNNs) with binary representations for synapse weights, input and output activations, as
these precisions offer the lowest hardware cost.
We also leverage pruning techniques to furthermore reduce the number of
operations in several networks, while having little impact on accuracy.

The key computational advantages obtained by using quantized arithmetic in inference are threefold:
Firstly, the quantized weights and activations have a significantly
lower memory footprint and the working set of some quantized neural networks (QNNs) may
entirely fit into on-chip memory. Fewer off-chip memory
accesses also mean orders of magnitude less energy
consumption. Furthermore, as on-chip memory can deliver much higher
bandwidth, the utilization of the compute resources is increased for a
higher performance.
For example, binarization reduces the model size
of VGG-16 \cite{DBLP:journals/corr/SimonyanZ14a} from 4.4\,Gbit to 138\,Mbit and that of
YOLOv2 \cite{redmon:2016} from 1.6\,Gbit to 50\,Mbit.
Secondly, replacing floating-point with fixed-point representations
inherently reduces processing power by orders of
magnitude \cite{horowitz20141}. While the reported study is for 45\,nm ASICs,
FPGAs follow similar general trends. Finally, the hardware resource cost of quantized operators is significantly smaller than that of floating-point ones. This allows a
much higher compute density with the same amount of resources and
thereby an easier performance scaling, as the key computation in neural network
inference is repeated multiply-accumulate (MAC) operations.
As will be shown in \rsec{secQNNimpl}, hardware complexity of an integer MAC basically grows
proportional to the bitwidths of both factors.

\begin{figure}
\centering
\includegraphics[width=0.8\linewidth]{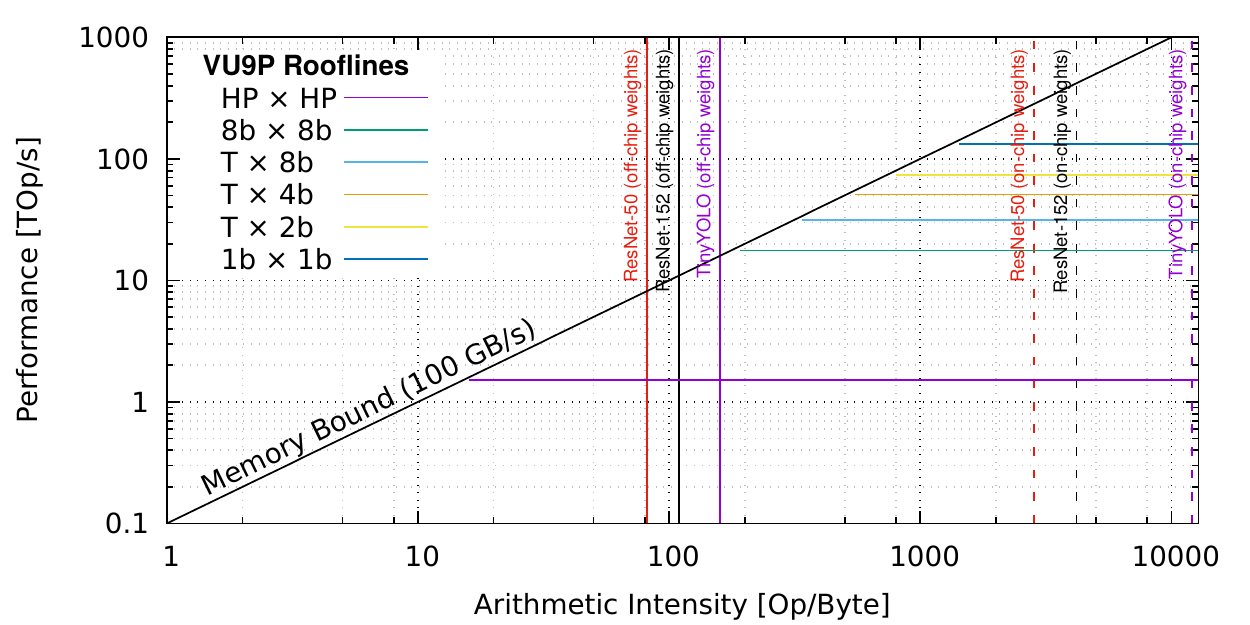}
\caption{AWS\,F1 Rooflines across a range of precisions.}
\label{fig:roofline}
\end{figure}
We use a roofline model, as introduced by Williams
et\,al. \cite{rooflines}, to exemplify the computational benefits of QNNs.
\rfig{fig:roofline} shows the peak performance of an AWS\,F1 FPGA node
for operations at different precisions\footnote{The following
assumptions were applied: clock frequency: 400\,MHz, 90\% DSP
and 70\% LUT utilization; HLS overhead included by hardware cost functions as
derived later.}.
The graphs illustrate that by going from half-precision floating point to
reduced-precision fixed point, the peak compute
performance increases by 87$\times$. We also show how the accompanying
reduction in model size enables staying in on-chip memory.
This is illustrated by the arithmetic intensity of three different neural network
topologies, specifically ResNet-50, ResNet-152 and Tiny\,YOLO, which are shown
for the two cases of the half-precision floating-point and 1-bit
operation.
The arithmetic intensity for the 1-bit variants is greatly increased,
and the implementation is no longer memory bound.

\begin{figure}
\centering
\includegraphics[width=0.7\linewidth]{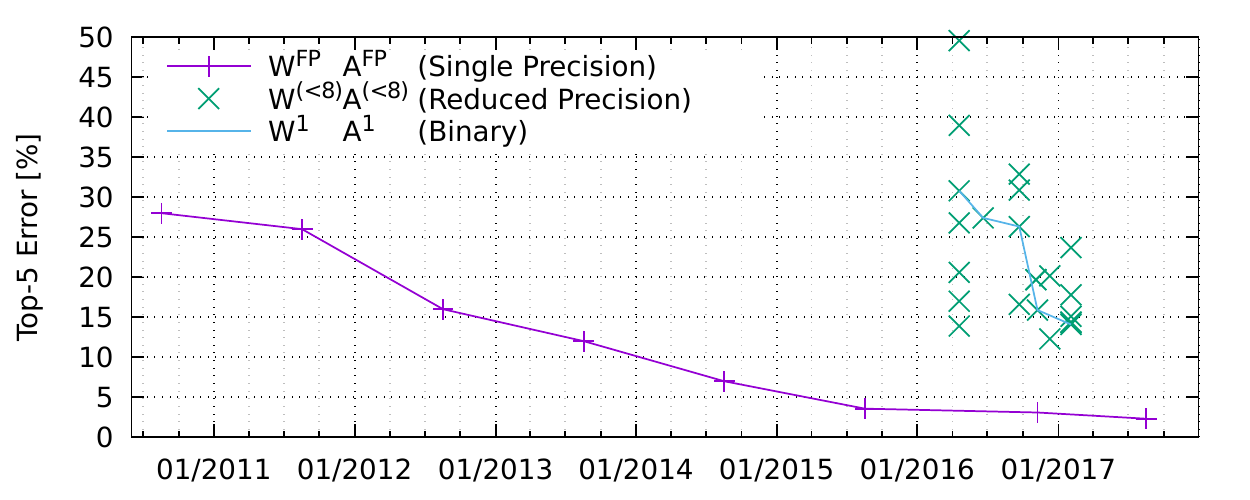}
\caption{Accuracy over time for ImageNet Classification.}
\label{fig:accovertime}
\end{figure}

The reduction in precision has a slight impact on
accuracy, which has been addressed by quanti\-zation-aware training
techniques, by innovative numerical representations
(e.g. MS-FP8 by Microsoft Research \cite{customfp}) and by
new quantization schemes (e.g. Half-wave Gaussian \cite{cai:2017}).
This is illustrated by \rfig{fig:accovertime} showing the published top-5 error rate
for the ImageNet classification by 32-bit floating-point networks \cite{ilsvrc} and
corresponding reduced-precision networks \cite{DBLP:journals/corr/RastegariORF16,cai:2017,DBLP:journals/corr/ZhouNZWWZ16,DBLP:journals/corr/HanMD15, iandola2016squeezenet, wideresnet,DBLP:journals/corr/ZhuHMD16}.

\begin{figure}
\centering
\includegraphics[width=0.7\linewidth]{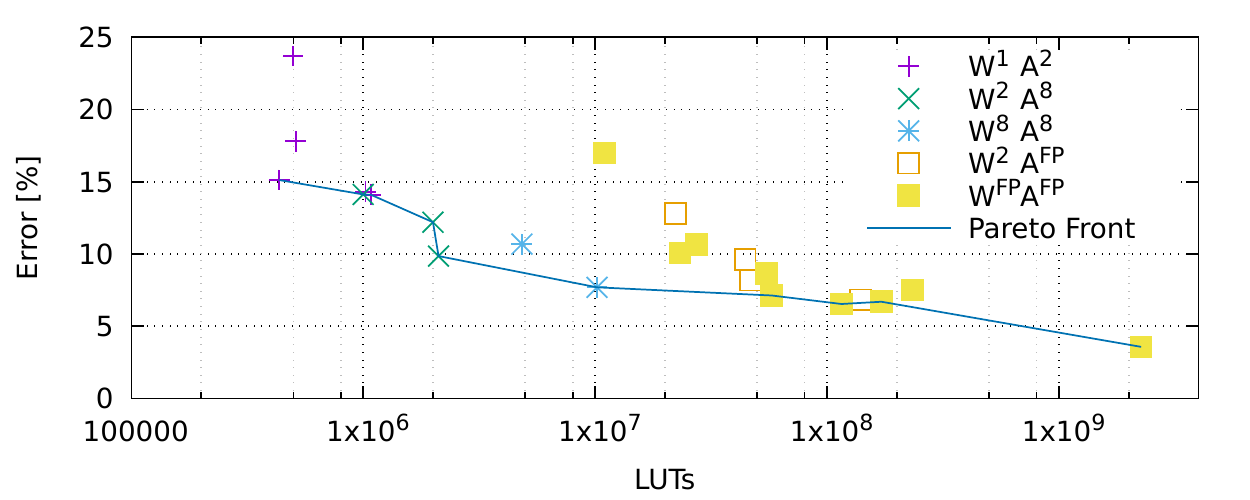}
\caption{Accuracy vs. Hardware Cost with Different Precisions for ImageNet.}
\label{fig:accuracy_cost}
\end{figure}

The minimal reduction in accuracy
combined with significant performance gains and power savings
provides highly interesting design trade-offs.
\rfig{fig:accuracy_cost} illustrates achievable compromises
for the selection of CNNs given above, depicting the relationship
between top-5 error rate and hardware cost for ImageNet classification.
For this, we assume a target frame rate of 10,000\,frames per second
(fps), clock rate of 300\,MHz, and a hardware cost in lookup tables (LUTs)
as derived by the microbenchmarks in \rsec{secQNNimpl} with HLS. The interesting points within this design spectrum are the
ones on the Pareto frontier. It is clear that, for example, for a
maximum error of $10$\%, the most cost-efficient
implementation leverages a 2b/8b representation. Similarly, these graphs illustrate
that Pareto optimal trade-offs are often reduced-precision networks, for instance when
the hardware cost or energy budgets are fixed.
This is the case in many
applications, be it a maximum price target for an embedded device or the PCIe power budget of 75\,W.

The key question is, given a set of design constraints and a specific
machine learning task, how the best possible trade-off within the vast
design space can be identified. This question entails two parts: One is concerned with deriving the most implementation-friendly neural network, and the second part looks at the hardware implementation itself.
Deriving systematic accuracy results is a time-intensive process given typical neural network training
times. But even the potential computational benefits cannot be
easily understood because hardware implementations are
time-intensive, the deployment space is complex, architectural choices
are myriad and the prediction of power, performance and latency is
complex. To find an optimal implementation given
a set of design constraints requires a framework that
provides insights and estimates given a set of design choices and
automates the customization of the hardware implementation.

\begin{figure}
\centering
\includegraphics[width=0.9\linewidth]{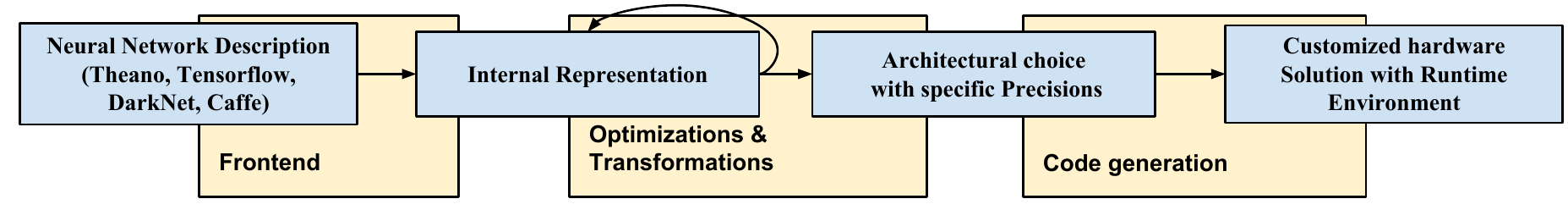}
\caption{\finnr Framework Overview}
\label{fig:finnintro}
\end{figure}

To address
this challenge, we implemented \finnr, the second version of the original
tool~\cite{umuroglu2017finn}, which supports more architectural choices as well
as mixed and variable precisions beyond binary. \finnr uses a
quantization-aware intermediate representation to enable QNN-specific
optimizations and has a modular frontend/transform/backend structure
for flexibility as is shown in \rfig{fig:finnintro}.
The focus of this article is on the architecture of inference accelerators, their optimization and automated generation towards different design targets.
While we discuss some techniques used towards the quantization of neural networks during training, this is currently still a highly active research area and well beyond the scope of this paper.
Our contributions are as follows:
\begin{itemize}
  \item Review and summary of the state of the art in reduced-precision neural networks, accuracy, frameworks and reconfigurable hardware accelerators.
  \item Microbenchmarks demonstrating performance-cost tradeoffs of different compute architectures and substrates for a broad range of precisions.
  \item Cost models for different architectures and precisions.
  \item A modular quantization-aware end-to-end framework for QNN exploration and implementation.
  \item Experimental results on four state-of-art QNN
    implementations on three different platforms yielding
    unprecedented measured peak performance.
\end{itemize}

This article is structured as follows: \rsec{secBackground}
reviews the state of the art in
reduced precision neural networks. We describe the hardware
architecture choices in greater detail in
\rsec{secQNNimpl}, including the microbenchmark results which are used to derive hardware
cost estimation functions, while \rsec{secFINN} contains the details on
the \finnr framework.
\rsec{secShowCases} presents experimental results, demonstrating the benefits of reduced precision and validating the \finnr workflow and its flexibility with results
measured for a range of neural networks, platforms and precisions.
Finally \rsec{secCon} concludes the article and provides an outlook to future work.

\section{Background}\label{secBackground}
QNN accuracy, hardware inference accelerators and frameworks to
automate the accelerator generation are fast moving fields of
research. The following sections capture the respective most recent
and relevant state of the art at the time of writing.

\subsection{Quantized Deep Neural Networks - Accuracy}
On smaller image classification benchmarks such
as MNIST, SVHN and CIFAR-10, QNNs have been demonstrated
\cite{courbariaux:2016,DBLP:journals/corr/ZhouNZWWZ16} to achieve nearly state-of-the-art accuracy.
Kim~and~Smaragdis~\cite{kim:2016} consider full binarization (where weights, inputs and outputs are binarized) with a predetermined portion of the synapses having zero weight, and all other synapses with a weight of one.
They report 98.7\% accuracy with fully-connected networks on the MNIST dataset, and observe that only XNOR and bitcount operations are necessary for computing with such neural networks.
XNOR-Net by Rastegari~et\,al.~\cite{DBLP:journals/corr/RastegariORF16} applies convolutional BNNs on the ImageNet dataset with topologies inspired by AlexNet, ResNet and GoogLeNet, reporting top-1 accuracies of up to 51.2\% for full binarization and 65.5\% for partial binarization (where only part of the components are binarized).
DoReFa-Net by Zhou~et\,al.~\cite{DBLP:journals/corr/ZhouNZWWZ16} explores reduced precision during the forward pass as well as the backward pass. 
Their results include configurations with partial and full binarization on the SVHN and ImageNet datasets,
including best-case ImageNet top-1 accuracies of 43\% for full and 53\%  for partial binarization.
For the more challenging ImageNet benchmark, there is a noticable accuracy drop when using QNNs compared to their floating point equivalents,
however
there is significant evidence that increasing network layer size can compensate for this drop in accuracy as shown by by Fraser et\,al. \cite{Fraser:2017:SBN:3029580.3029586}, Sung et\,al. \cite{sung:2015}, Zagoruyko et\,al. \cite{wideresnet}, Mishra~et\,al.~\cite{DBLP:journals/corr/abs-1709-01134} and Kim et\,al. \cite{kim:2016}.

Furthermore, new quantization schemes show promising results.
For instance, Cai et\,al. \cite{cai:2017} proposed
Half-wave Gaussian Quantization (HWGQ) to take advantage of the
Gaussian-like distribution of batch-normalized activations,
demonstrating QNNs with binary weights and 2-bit activations with less than 5\% top-5 accuracy
drop compared to floating point DNNs on the challenging ImageNet
dataset, as summarized in \rtab{tabHWGQaccuracy}. To the best of our knowledge, currently lowest error rates for ImageNet classification have been achieved using ternarization \cite{alemdar2016ternary, DBLP:journals/corr/ZhuHMD16}.

\begin{table}
  \caption{Latest Accuracy of Several QNNs \cite{cai:2017,alemdar2016ternary, DBLP:journals/corr/ZhuHMD16} on ImageNet dataset}
  \label{tabHWGQaccuracy}
  \resizebox{0.4\linewidth}{!}{\begin{tabular}{ccc}\toprule
    Network & float top-1(top-5) & QNN top-1(top-5)\\\midrule
    GoogLeNet & 71.4\% (90.5\%) & 63.0\% (84.9\%) \\
    VGG-like & 69.8\% (89.3\%) & 64.1\% (85.6\%) \\
    ResNet-50 & 79.26 (94.75\%) & 64.6\% (85.9\%) \\
    \bottomrule
  \end{tabular}}
  \vspace*{-3ex}
\end{table}

While different numerical representations are worth investigation,
our current focus are quantized values with fixed integer representations below 8\,bits.
We use \emph{integer} to also refer to fixed-point numbers as we can absorb
fixed-point scaling factors into thresholds. The following notation \qnn{x}{y} is used across this article to represent
a layer with $x$-bit weights and $y$-bit activations.
The QNN networks within this article have most layers heavily quantized but
may contain higher-precision or even floating-point layers. As pointed out before, the discussion on how to train for reduced precision is
referenced in the experimental section for each QNN.



\subsection{Accelerators \& Architectures}
A great deal of prior work on mapping neural networks to hardware exists for both FPGAs and ASICs.
We refer the reader to the work by Misra and Saha~\cite{surveyannhw} for a comprehensive survey.
We cover a recent and representative set of works here, roughly dividing them into three categories based on their basic architecture:
1) a single processing engine~\cite{ovtcharov2015accelerating,zhang2015optimizing,chen2016eyeriss,andri2016yodann,chen2016diannao,judd2016stripes,jouppi2017datacenter,moss2017high,liang2017fp}, usually in the form of a systolic array, which processes each layer sequentially;
2) a streaming architecture~\cite{venieris2016fpgaconvnet,alemdar2016ternary,prost2017scalable}, consisting of one processing engine per network layer where inputs are streamed through the dataflow architecture and all layers are computed in parallel;
3) a vector processor~\cite{farabet2009cnp} with instructions specific to accelerating the primitive operations of convolutions.
Recent graphics processing units (GPUs) have been shown to deliver competitive results, 
as well as neurosynaptic processors, which implement many digital neurons and their interconnecting weights \cite{esser2016convolutional}. However this study focuses on FPGA based accelerators only.
For a detailed performance and efficiency comparison, we refer the reader to \rsec{secShowCases} and particularly \rtab{tabImplementations}.

\emph{Single Processing Engines:} Zhang~et\,al.~\cite{zhang2015optimizing} describe a systolic array style architecture, using theoretical roofline models to design accelerators optimized for the execution of each layer.
Ovtcharov et\,al. \cite{ovtcharov2015accelerating} implement a similar style architecture achieving a 3$\times$ speedup.
Eyeriss by Chen~et\,al.~\cite{chen2016eyeriss} use 16-bit fixed point rather than floating point, and combine several different data reuse strategies which
provide 2.5$\times$ better energy efficiency over other methods.
YodaNN by Andri~et\,al.~\cite{andri2016yodann} have a similar design as Zhang~et\,al.~\cite{zhang2015optimizing} but explore binary weights for fixed sized windows.
Moss~et\,al.~\cite{moss2017high} also implement a systolic array style processor, but specifically for BNNs, allowing for very high throughput, up to 40.8\,TOp/s.
Some alternative approaches to single accelerator designs are:
Stripes~\cite{judd2016stripes}, which implements a bit-serial processor capable of handling multiple precisions on a single compute array.
The authors experiment with precision from 3 to 13 bits.
FP-BNN, another implementation of a single processing engine for BNNs, utilizing an XNOR-popcount datapath. Interestingly, the authors implement batch-normalization and scaling in floating point for some networks, resulting in higher DSP usage than perhaps is required.

\emph{Streaming architectures:} Venieris and Bouganis~\cite{venieris2016fpgaconvnet} proposed a synchronous dataflow (SDF) model for mapping CNNs to FPGAs, which is a similar approach to our dataflow variant. 
Their designs achieve up to 1.62$\times$ the performance density of hand tuned designs.
Alemdar~et\,al.~\cite{alemdar2016ternary} implement fully-connected ternary-weight neural networks with streaming and report up to 255K frames per second on the MNIST dataset, but concentrate on the training aspect for those networks. Baskin~et\,al.~\cite{baskin2017streaming} similarly map multibit CNNs in streaming fashion onto FPGAs and show superior performance over GPUs.
Prost~et\,al.~\cite{prost2017scalable} design ternary networks in a dataflow, achieving notably high accuracies and performance, likely due to ternarization and a hand optimized RTL design.


\emph{Vector processors:} Farabet~et\,al.~\cite{farabet2009cnp} describe a programmable ConvNet Processor (CNP), which is a RISC vector processor with specific macro-instructions for CNNs including 2D convolutions, 2D spatial pooling, dot product and an elementwise non-linear mapping function.
The authors also created a tool to compile a network description into host code which is used to call the CNP.




\emph{Accelerator frameworks:}
Numerous new frameworks, including the original FINN tool~\cite{umuroglu2017finn}, have been proposed that take a graph based description of neural networks (such as Caffe's prototxt format) to operational hardware implementations, whereby some of those leverage a fixed hardware architecture, and others customize the hardware accelerator to achieve better throughput, latency or power reduction.
To the best of our knowledge \finnr is the only tool that supports arbitrary precision in weights, input and output activations, plus the flexibility in the backend which includes two hardware architectures to support a spectrum of design goals, as well as multiple target platforms.
\emph{DNNWeaver}~\cite{sharma2016dnnweaver} is a tool which generates bitstream+host code implementing CNNs on several FPGA platforms on the basis of a Caffe prototxt description. The generated coprocessor implements multiple processing engines depending on available resources. External memory is used for weights and for intermediate feature maps while the arithmetic supports 16-bit fixed or floating point. 
Similarly, \emph{fpgaConvNet} is not quantization-aware. It creates
customized FPGA dataflow architectures using reconfiguration when
designs do not fit \cite{venieris2016fpgaconvnet}.
\emph{CaffePresso} focuses on embedded systems with 20\,W power budgets including the Xilinx ZC706 (FPGA), NVIDIA Jetson TX1 (GPU), TI Keystone II (DSP), and Adapteva Parallella (RISC+NoC). The tool is currently limited to low-complexity classifiers which operate on small image maps and few class labels. Combining auto-tuning of the implementation parameters, and platform-specific constraints deliver optimized solutions for each input ConvNet specification. 
\emph{GUINNESS}~\cite{nakahara2017demonstration} is a GUI-based tool
flow for training BNNs on GPUs and deploying them using Xilinx devices
through SDSoC.
Similarly, \emph{HADDOC2} \cite{abdelouahab2017tactics}, the synthesis tool described by Wei~et\,al.~\cite{wei2017automated} and the compiler proposed by Ma~et\,al.~\cite{ma2017automatic} transform CNN descriptions to synthesizable hardware.
Finally, \emph{Minerva} \cite{reagen2016minerva} proposed a 5-stage SW-HW co-design work flow of an inference engine, however is constrained to fully-connected (FC) layers only. Unlike \finnr, Minerva includes a training space exploration, which is used to explore accuracy / resource trade-offs with a layer-wise direct quantization scheme, synapse pruning using thresholding and SRAM fault mitigation.

\section{Inference Accelerator Architecture}\label{secQNNimpl}
In this section, we investigate the various possible architectural choices when mapping inference onto programmable logic.
The first subsection takes an in-depth look at reduced precision
operational costs for both LUT- and DSP-based implementations. This
includes systematic benchmarking results for a broad choice of
precisions, which we use as the basis for our \emph{operation cost function}. In the second
subsection, we discuss the implementation of individual layers,
including their related \emph{layer cost function}. The third
subsection elaborates on the supported choices for a complete
inference accelerator and their associated impact on the resource
requirements yielding the \emph{accelerator cost function}.
The cost functions are essential to obtain the optimal levels of parallelism for the architecture and performance predictions.

\subsection{Microbenchmarks and Operation Cost Function}
\label{sec:ubenchmark}
For a thorough understanding of the resource requirements of the
omnipresent dot product computation, we have designed and implemented a set of
microbenchmarks which perform multiple multiply-accumulate operations
that can be customized in the bit widths of both operand vectors and
the number of products summed up in a single step. We have
implemented these microbenchmarks both (a) in VHDL for traditional
RTL synthesis and (b) in C++ targeting the HLS flow.
For \finnr, we have chosen the latter for a number of reasons, namely
design productivity, portability to different
platforms, built-in optimizations for pipelining, design space exploration
and automated flow control. However this comes at the expense of some resource overhead.
The RTL results allow us to estimate the overhead for the higher-level design entry and establish a reference for
potential future gains.


Combinations of different dot product sizes, which is defined as the size of the input vectors, $N \in\{1,2,3,4,6,8,12,16,24,32,48,64,96,128\}$ and
operand bit widths $W\times A \in\{1,2,3\}\times\{1,2,3\}$ were measured, resulting in a
three-dimensional parameter space.
An elementary multiply decomposes into (a) the generation of
a skewed bit matrix produced from $W\cdot A$ AND operations and (b) its
summation to yield the value of the product. The structural
combinational complexity of both steps is determined by the number of
bits produced and added up, respectively. The LUT cost can
thus be expected to grow proportionally to this product as well, give
or take some jitter caused by functional fragmentation due to the
required mapping to physical 6-input LUTs.

Scaling the size $N$ of the dot product to longer vectors can be
expected to scale the structural complexity of the computation
accordingly. Note that $N$ bit matrices of size
$W\cdot A$ must be produced. The following summation would ideally
operate on all the stacked and merged matrices together without
producing the individual partial results. This additive reduction
would have $N\cdot W\cdot A$ inputs suggesting a proportional
structural complexity in terms of LUTs.

The described implementation of the multi-MAC operation
is effectively enforced in our RTL implementation by employing the
generic matrix summation approach by Preu{\ss}er
\cite{preusser:2017}. The HLS flow through Vivado\,HLS and Vivado
implements a similar slightly less efficient adder tree after
the generation of the bit matrices. Both implementations
can be expected to induce structural LUT costs that are roughly
proportional to the product $C=N\cdot W\cdot A$, which we use to
describe the \emph{complexity} of the dot product operation. The
coefficients corresponding to the two implementation choices can be
determined empirically and is an immediate
measure of their relative efficiencies.

\begin{figure}
  \centerline{\includegraphics[width=.78\linewidth]{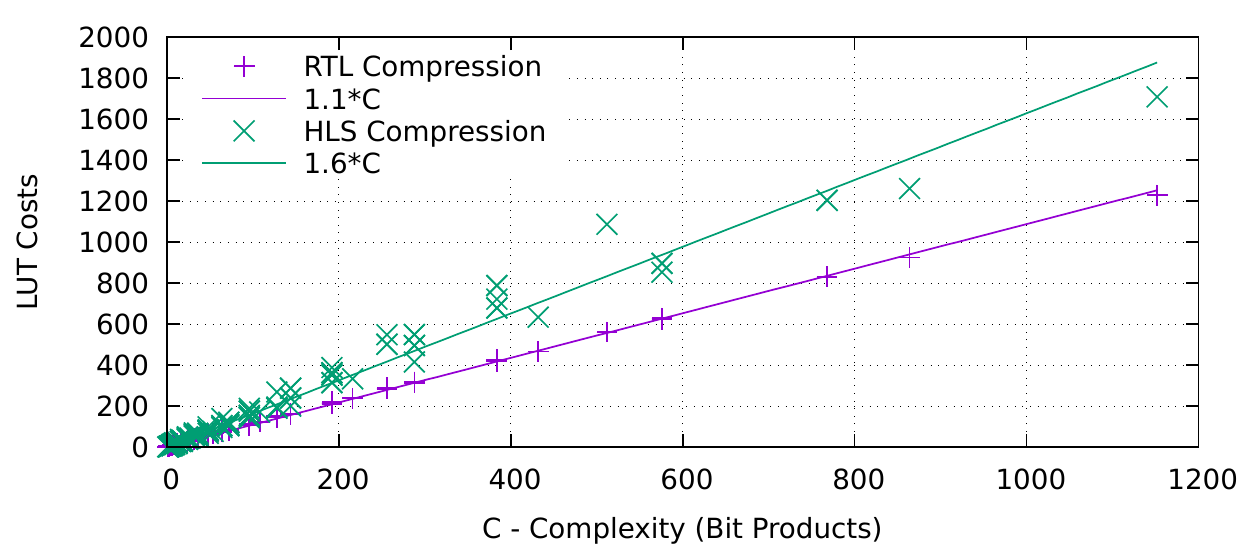}}
  \caption{LUT Costs of Dot Product Computation}
  \label{figMmacLuts}
\end{figure}

The results of our microbenchmark experiments are shown in
\rfig{figMmacLuts}. The anticipated linear dependency on the complexity
measured by the number of overall bit matrix elements is confirmed
to be extremely close for the RTL implementation while experiencing
a somewhat greater variation about the fitted center line for HLS
synthesis. HLS typically but unnecessarily reduces
the partial products completely to a conventional binary number before
feeding them into the adder tree. This creates an overhead
that grows with the size $N$ of the dot product and accounts for the
observed variation. Comparing the measured coefficients, the
HLS implementation is found to currently induce a 45\% resource
overhead over the far more complex VHDL implementation.

The use of binary ($\{-1,1\}$) and ternary ($\{-1,0,1\}$) weights is
very popular for reduced-precision neural networks. So, it is
interesting how these specific range types behave in comparison to the
neighboring conventional 1-bit or 2-bit integer types when used in
dot products with activations of various precisions. Factoring out $W$
from our previous complexity measure, we can predict linear
dependencies of the LUT costs on the remaining $C'=N\cdot A$
product. Using HLS only in these experiments, we find the
greatest cost increase going from conventional 1-bit weights, which
are rarely used, to binary weights. This is due to
the fact that already the negative multiple of $A$ now required in the
computation mandates the allocation of an extra sign bit. This can
only be mitigated in the trivial case of 1-bit activations using the
approach practiced by XNOR-Net or traditional binarized
FINN~\cite{umuroglu2017finn}. Otherwise, an increase in LUT costs of 35\% is
experienced. The additional increases of 20\% for each further step
going to ternary and 2-bit precisions are somewhat smaller.


\subsection{Layers}
\label{secHWLayers}
The prinicipal elements that compose a typical convolutional layer are the \emph{matrix-vector threshold
unit} (MVU) and the \emph{sliding window unit} (SWU).
MVUs handle the compute aspects: For convolutional layers, the convolutions themselves can be
\emph{lowered}~ to matrix-matrix multiplications, which is well understood \cite{chellapilla2006high}.
These can then be mapped in a streaming fashion
onto the MVU. The corresponding weights from the convolution filters are packed into a
filter matrix, while a sliding window is moved across input images
to form an image matrix. These matrices are then multiplied to generate the output images.

\begin{figure}
  \begin{center}
  \includegraphics[width=0.3\linewidth]{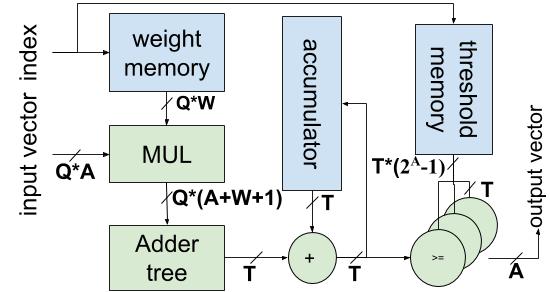}
  \end{center}
  \caption{Processing Element (PE) as Basic Compute Component}
  \label{figPE}
\end{figure}
\begin{figure}
    \begin{subfigure}{.55\linewidth}
        \includegraphics[width=\linewidth,trim=0 70 0 1,clip]{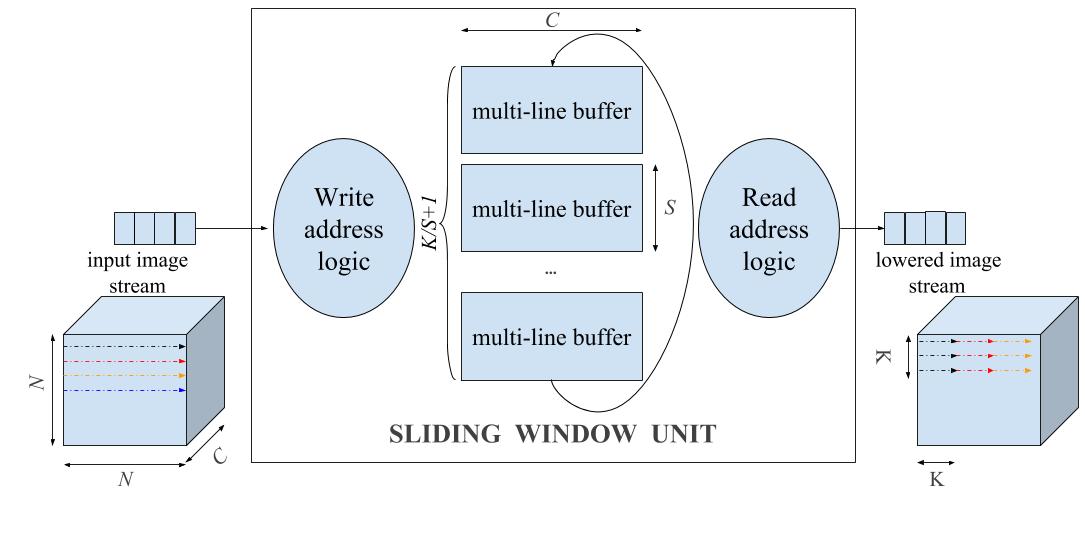}
        \caption{Sliding Window Unit (SWU)}
        \label{figSWU}
    \end{subfigure}\hfill
    \begin{subfigure}{.38\linewidth}
        \includegraphics[width=\linewidth,trim=0 0 0 9,clip]{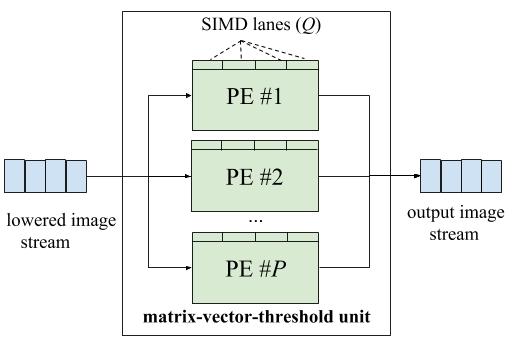}
        \caption{Matrix-Vector-Threshold Unit (MVU)}
        \label{figMVTU}
    \end{subfigure}
    \caption{SWU and MVU Block Diagram}
\end{figure}

We refer to the principal compute component of convolutional or fully
connected layers as a \emph{processing element} (PE). Its structure is
illustrated in \rfig{figPE}. A PE performs $Q$ parallel
multiplications, which corresponds to the SIMD value. It then reduces
them in an adder tree for their subsequent accumulation towards the
currently computed dot product. Finally, threshold comparisons are
used to derive the output values from the accumulation results. An
array of $P$ parallel PEs comprises a MVU.
A third degree of concurrency is introduced which supports computation of multiple
output pixels sharing the same weights within the same channel in parallel, referred to as $M$. This enables performance scaling with increased BRAM utilization.
The choices of the parameters $P$, $Q$ and $M$ determine the degree of a layer's computational
parallelism. They are the key parameters for trading off resource versus performance of any layer's computation.

The SWU is the unit that generates the image representation required for a
convolution lowered to a matrix multiplication (\rfig{figSWU}).
It generates the same vectors as those in~\cite{chellapilla2006high} but with
interleaved channels~\cite{umuroglu2017finn} to simplify memory accesses and to avoid the need for transposition between layers.
This exhibits significantly lower latency compared to full image buffers
and reduces buffer size requirements. Only as many consecutive rows as
the height of the convolutional kernel must be kept available.
For elasticity reasons, an extra row is used to collect
new incoming image data.

%

\subsubsection{Layer Cost Model}
\begin{figure}
  \centerline{\includegraphics[width=\linewidth]{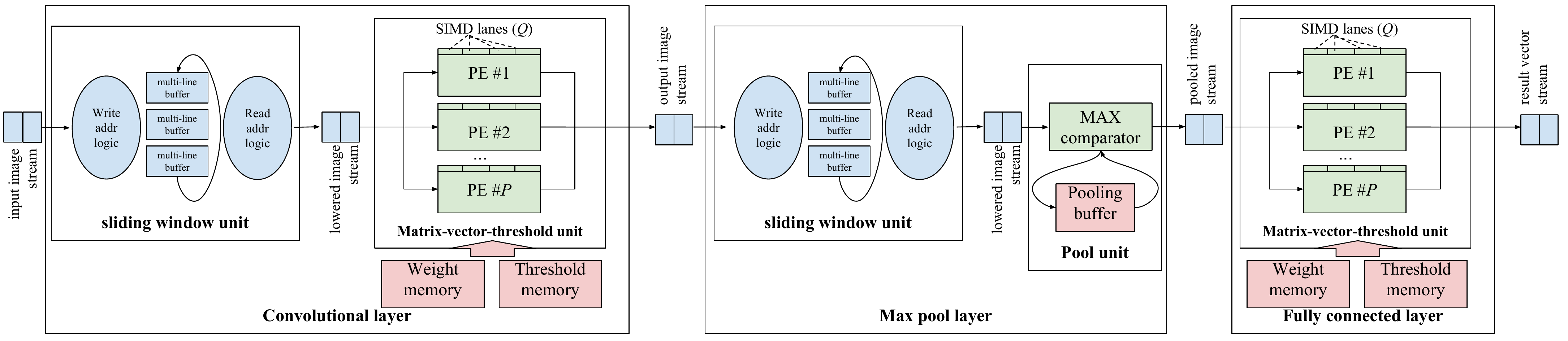}}
  \caption{Possible Layer Compositions}
  \label{figLayers}
\end{figure}

As can be seen from \rfig{figLayers}, a layer is composed of
different elements. Convolutional layers are composed of SWU, MVU as
well as weight and threshold memories (WM \& TM). Maxpool layers
contain a SWU and a maxpool unit, and fully connected layers require
only MVUs. Thus, the layer cost is a sum of the basic components as
shown in \req{eqLayer}.
Note that the logic cost relating to WM ($LUT_{\mbox{WM}}$) and the BRAM cost of MVUs ($BRAM_{\mbox{MVU}}$) is neglible.

\begin{align}
 \label{eqLayer}
BRAM_{\mbox{CNV}} &= BRAM_{\mbox{SWU}} + BRAM_{\mbox{WM}};
&LUT_{\mbox{CNV}} &= LUT_{\mbox{SWU}} + LUT_{\mbox{MVU}}\\
BRAM_{\mbox{FC}} &= BRAM_{\mbox{WM}};
&LUT_{\mbox{FC}} &= LUT_{\mbox{MVU}} \\
BRAM_{\mbox{MP}} &= BRAM_{\mbox{SWU}} + BRAM_{\mbox{MP}};
&LUT_{\mbox{MP}} &= LUT_{\mbox{SWU}} + LUT_{\mbox{MP}}
\end{align}

In the following paragraphs, we derive BRAM, LUT and DSP costs for all components (SWU, MVU, WM, MP) separately.

\begin{table}
\caption{Layer Parameters}\label{tabParams}
\resizebox{0.8\textwidth}{!}{
\centerline{\begin{tabular}{@{\quad}lll}\toprule
\textbf{Fully Connected Layers} & $D,D'$             & input \& output vector size\\
\textbf{Convolutional Layers} & $N, C$         & input feature map width and channels\\
   &$K\times K, S$ & kernel dimension and stride\\
   &$C'$           & output feature map channels\\
\textbf{MVU Dimensioning}   &$M$            & parallel vectors processed by MVU\\
   &$P, Q$         & number of output \& input channels computed in parallel\\
   &$A, W$         & bit width (precision) of activations / weights\\
\bottomrule
\end{tabular}}}
\vspace*{-3ex}
\end{table}

\paragraph{SWU Cost} The sliding window unit's hardware cost is dominated by the BRAM requirements which can be directly derived from
the implemented memory layout, as given by the parameters in \rtab{tabParams}. The line buffer occupies as many BRAM
modules as specified by \req{eqSWURAM}.
\begin{align}
  \label{eqSWURAM}
  \mbox{BRAM}_{\mbox{swu}} &= M\cdot\left(\left\lceil\frac{K}{S}\right\rceil+1\right)\cdot
  \left\{\left\lceil\frac{S\cdot N}{512}\right\rceil\times
  \left\lceil\frac{C\cdot A}{36}\right\rceil\right\}
\end{align}
The multi-vector count scales linearly at the highest level of the
equation. Otherwise, independent stripes of memory are used for each
set of rows that can be released independently once the whole width of
a line has been processed. An additional memory stripe is used as
assembly buffer for the new image data coming in. This accounts for
the first parenthesized factor. The remaining two factors capture the
depth and the width of the memory stripes, which are potentially
fragmented due to the depth and word width of the built-in BRAM
modules.
There is also a constant overhead in logic resources. This varies depending on the type of accelerator architecture.
For a full feed-forward dataflow, each SWU requires 426\,LUTs and 0\,DSPs as the exact dimensions are known at compile-time and the parameters can be baked into the architecture.
For a multilayer offload that executes many different layers on top of the same hardware components, the parameterization happens at run-time, therefore the overhead is larger
with 1050\,LUTs and 15\,DSPs respectively.


\paragraph{WM Cost}
For convolutional layers, \req{eqWEIGHTRAM} captures the number of BRAM
modules needed to implement the weight memory of a convolutional
layer. Its overall size is determined by the product of the squared
kernel dimension and the numbers of input as well as output feature
map channels. This memory volume is split into separate memories, one
for each processing element. The parallel access of $Q$ weights
determines the word width used by the implementation. Again, memory
depth and word size may be fragmented by the physical dimensions of
the available BRAM modules.
\begin{align}
  \label{eqWEIGHTRAM}
  \mbox{BRAM}_{\mbox{WM}} &= P\cdot\left\{\left\lceil\frac{\omega}{512}\right\rceil\times
  \left\lceil\frac{Q\cdot W}{36}\right\rceil\right\}
  &\mbox{with}\quad \omega = \left\{\begin{array}{cl}
  \frac{K^2\cdot C \cdot C'}{Q\cdot P} & \mbox{\footnotesize for convolutional layers}\\
  \frac{D \cdot D'}{Q\cdot P}          & \mbox{\footnotesize for fully-connected layers}
  \end{array}\right.
\end{align}

\paragraph{MVU Cost}
The computational concurrency of a convolutional
layer is controlled by (a) the number $P$ of PEs
concurrently working on distinct output channels,
(b) $Q$, the SIMD of input channels processed
within one clock cycle, and (c) $M$, the multi-vector count capturing the
concurrent duplication of this compute structure across multiple output
pixels for convolutional layers. These parameters allow to scale the
performance of a layer implementation in a wide range but also affect
the hardware costs directly. Generating a network implementation, \finnr
must be aware of these costs in order to be able to scale the
individual layer implementations towards a balanced
performance within the resource limits of the
targeted device.

The hardware cost of the MVU can be modeled as an essentially constant
control part and the dot product arithmetic. The latter scales both with
the duplication into parallel PEs and with parallel multi-vector
processing. The model for the internal costs of the individual
arithmetic blocks can be taken from the results of the microbenchmarks
in \rsec{sec:ubenchmark}. Combining both control and
arithmetic, we derive the following model:
$LUTs = c_0 + c_1\cdot M\cdot\left(P\cdot Q\right)\left(W\cdot A\right)$.
Recall that $M=1$ for fully connected layers as they cannot share
weights across multiple kernel applications.

\begin{figure}
  \centerline{\includegraphics[width=.78\linewidth]{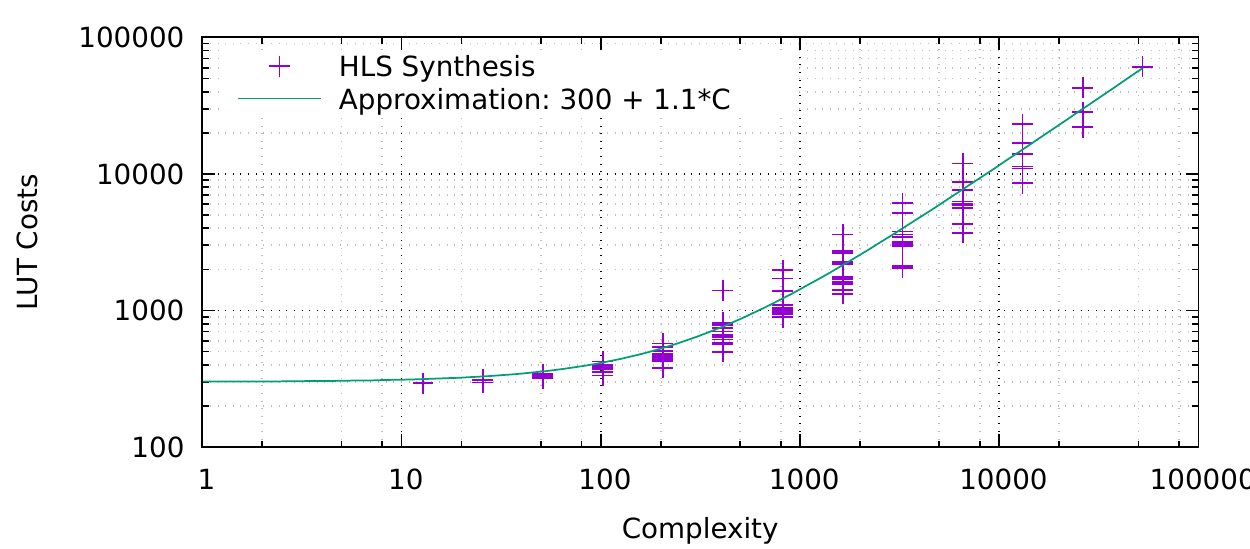}}
  \caption{Empirical Fit of the LUT Cost Model for the MVU}
  \label{figMVUmodel}
\end{figure}
To determine the two parameters of this model and to validate its
fitness, we again performed HLS synthesis experiments with the
parallelization parameters $M=1$, $P\in\{2,4,8,16,32,64\}$ and
$Q\in\{2,4,8,16,32,64\}$. The complete product scaled by $c_1$ is
taken as the single-figure complexity measure. The obtained
empirical fit of the LUT model against the synthesis results is
depicted in \rfig{figMVUmodel}. The anticipated behavior is confirmed;
however, the prediction may be off by up to
30\% of the later synthesis result for individual experiments.

The cost of the threshold operations depends strongly on the
precision of the output activation function as the number of
thresholds to be stored and compared with grows exponentially with
this precision. For small precisions like the ones \finnr is targeting,
these costs practically disappear within the remaining MVU
costs. Going for precisions significantly above 4~bits will quickly
render the associated costs expensive or simply make the thresholding
approach infeasible altogether.

\paragraph{MP Cost}
The BRAM and LUT requirements for the actual compute of the max
pooling layers is very little. The block is basically implementing $C$
parallel comparators, one for each channel whereby each sequentially compares
two $A$-bit words holding onto the maximum of its pooling window. The
total computational LUT costs are roughly equivalent to the product of $A$ and $C$.

\subsection{Full Inference Accelerator Architecture: Dataflow and Multilayer Offload}
\label{faa}
\begin{figure}
    \begin{subfigure}{.52\linewidth}
	\centerline{\includegraphics[width=\linewidth]{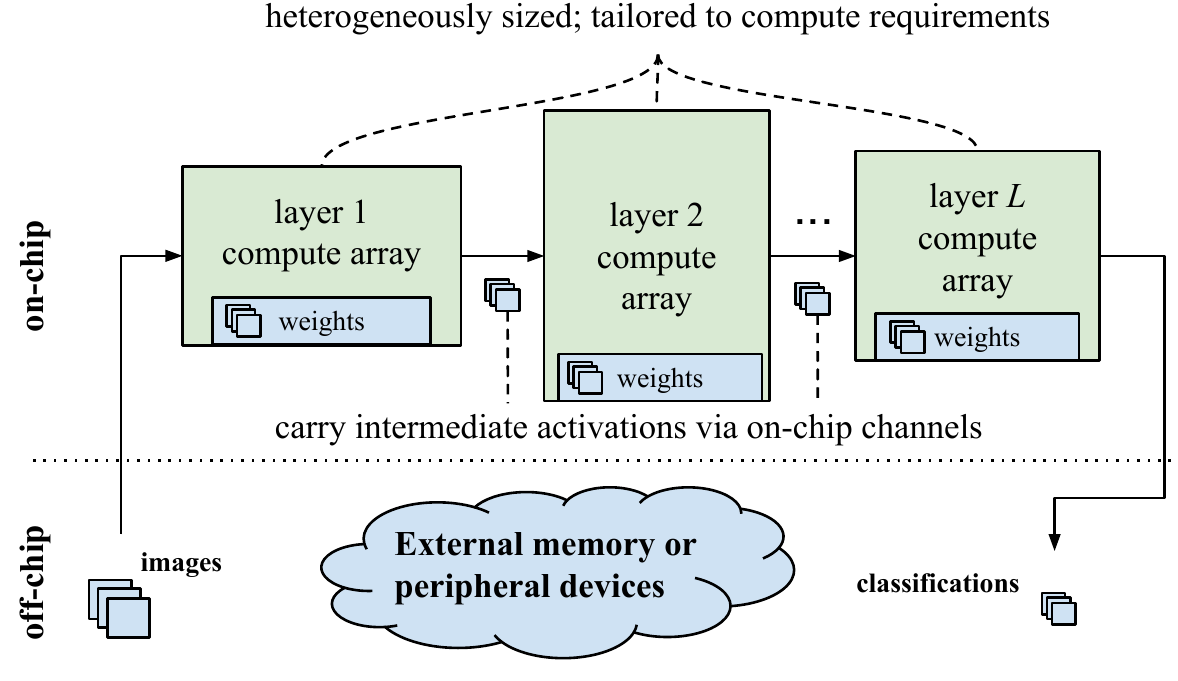}}
	\caption{Dataflow Architecture with Per-Layer Tailored Compute Arrays, On-Chip Weights and Activations}
	\label{figPipeline}
    \end{subfigure}\hfill
    \begin{subfigure}{.4\linewidth}
	\centerline{\includegraphics[width=\linewidth]{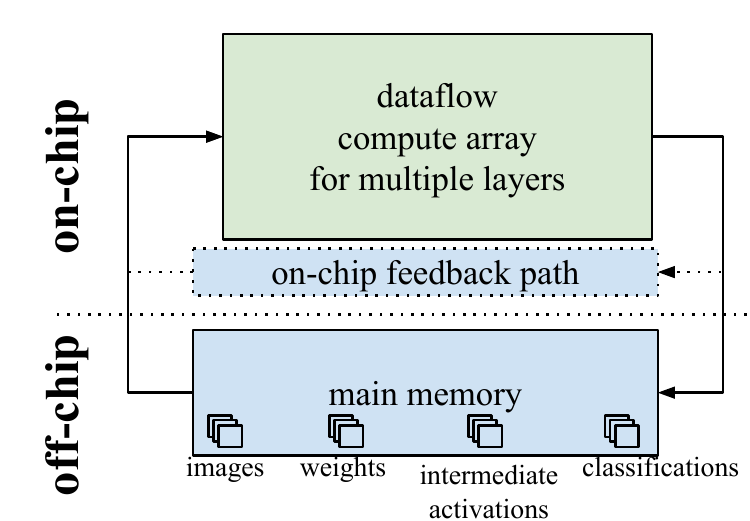}}
	\caption{Multilayer Offload Architecture with Maximally-Sized Homogeneous Compute Arrays for Different Precisions}
	\label{figLoopback}
    \end{subfigure}
    \caption{Possible Backend Architectures}
\end{figure}

\finnr supports two key choices for the architecture of the accelerator.
The first is a custom-tailored strictly feed-forward dataflow implementation as described in the original FINN paper \cite{umuroglu2017finn} and illustrated in \rfig{figPipeline}.
The second offloads a part of the dataflow pipeline which represents a significant proportion of the compute load and allows the feature maps to iterate over it multiple times through a loopback path as is shown in \rfig{figLoopback}.

The customized \emph{Dataflow Architecture (DF)} differentiates itself from many other accelerators in that it is customized for a specific NN topology and for different precisions in activations and weights for each individual layer avoiding \enquote{one-size-fits-all} inefficiencies and reap more of the benefits of reconfigurable computing. One streaming compute engine is instantiated per layer, with resources tailored to fit each layer's compute requirements and the user-defined frame rate. This is accomplished by adjusting their $P$, $Q$ and $M$, as introduced in \rsec{secHWLayers}, according to the algorithm in \rsec{secDataflowBalance}. As the datafow architecture is fully rate balanced, each layer produces and consumes data in the same order with the aim of minimizing buffer requirements in between layers and latency. An engine starts to compute as soon as the previous engine starts to produce output and as such introduces another level of concurrency between layers.

The \emph{Multilayer Offload Architecture (MO)} is beneficial
when the minimal footprint of the fully unrolled DF architecture
exceeds the target device capabilities or when the fragmentation
overhead is unattractive. The main application context are
large networks under hard resource constraints.

Comparing these two architectures, we observe the following:
The cost of the DF implementation is the sum
across all implemented layers, while the cost of the MO architecture is
defined by the maximum across the scheduled layers and provides as such better scalability towards really deep CNNs.
From a throughput point of view, which is dominated by the total amount of
compute resources instantiated and their utilization, we expect both
architectural choices to be equivalent. For DF, we experience a
certain amount of fragmentation as will be explained in the next
section, while for MO architectures, the utilization is determined by
how well a layer can be scheduled onto the offloaded compute
resources. However, we expect that the reduction of buffering between
layers should bring significant latency benefits for DF vs MO, but
this still remains to be confirmed with experiments.
Note that these two choices represent the two endpoints within a large
design space of potential architectures that could be worth a more
thorough investigation. For now, the chosen architectures provide
the flexibility to build accelerators that scale both to extreme
performance and to minimal footprints, even for very large networks.

\subsection{Full Accelerator Cost Model}
\label{facm}
The full accelerator cost encompasses a constant part given by the formal $shell$ resource overhead, and the dynamic aspect for the actual accelerator architecture, which is described above.
Depending on the chosen target platform, a different base infrastructure or $shell$ is created, which handles all memory interfaces, DMA engines, network connections and host interface.
\finnr currently supports a number of different platforms, whereby within the article we focus on PYNQ-Z1, AWS\,F1, and a Zynq UltraScale+ platform, called Ultra96. More detailed platform descriptions will be provided in \rsec{secShowCases}.
The shells are substantially different:
Ultra96 and Pynq-Z1 move the images and the results through the coherently shared memory between ARM and FPGA fabric (PS memory), and, as memory controllers are hardened inside the SOC, the corresponding soft logic requirements are very small (8\,BRAM18s, 2.6\,kLUTs).
For AWS\,F1, the SDK based design entry is chosen, which moves the data between host and FPGA card via FPGA-attached DRAM and requires soft memory controllers as well as PCIe interface with DMA engine, all joined with an AXI interconnect. The overall overhead amounts to 1090\,BRAM18s and 297\,kLUTs.
The total hardware cost for the different platforms can be computed as the sum of the specific shell overhead and the chosen accelerator architecture.



\section{\finnr}\label{secFINN}

\finnr is trying to answer the question: Given a set of design constraints and a specific
neural network, what is the best possible hardware implementation that can be achieved?
For this \finnr provides insights and estimates and
automates the customization of the hardware implementation.
The tool is to be used interactively to explore a given CNN in terms of high-level concepts of the target platform, architecture, and precisions to achieve specific design goals and satisfy given constraints.
Given the choices, the tool then customizes the hardware accelerator, either DF or MO style, to meet the constraints.
We currently support resource footprint and throughput constraints. Latency, power estimation, and automated design space exploration are left as future work.

The key functionality in the tool for the MO is the generation of the runtime schedule that sequences the compute onto the hardware engines, and
for DF, the calculation of the folding factors that generate a balanced dataflow whereby the whole architecture gets incrementally unfolded until design targets are met.
As shown in \rfig{fig:finnintro}, \finnr has a modular structure inspired by LLVM compiler infrastructure, with frontends, passes and backends, and a quantization-aware intermediate representation (IR) of QNNs. The \emph{frontend} is responsible for interfacing with a selection of training frameworks such as Caffe, DarkNet and Tensorflow and translating trained QNNs into the IR. The IR is used as the basis of the performance estimation tool. \finnr supports a number of \emph{transformations} that help generate more efficient representations. Finally, the \emph{backend} contains a code generator that creates executable inference accelerators for a selection of platforms, including PYNQ-Z1, Ultra96, and AWS\,F1. The accelerators are composed of components defined in the \emph{QNN library}. All of these components are described in further detail in the following subsections.



\subsection{Frontends}
The frontend stage is responsible for converting QNNs trained by a variety of frameworks to the \finnr intermediate representation. As each framework exposes their QNN topologies through custom formats, \finnr must first perform a conversion to a common intermediate representation, in order to process the network.
Currently, we support frontends for BinaryNet~\cite{courbariaux:2016}, Darknet~\cite{darknet} and Tensorpack~\cite{ZhouYGXC17}. In the case of BinaryNet or Tensorpack, \finnr examines the dimensions of these frameworks' network data stored in .npy files. Finally for Darknet, the network topology is extracted from the configuration files (.cfg files) of the network. As \finnr follows a modular design, additional frontends supporting new QNN frameworks can be added as they emerge.

Once converted to the intermediate representation, \finnr is aware of the network topology and the precision of the data types within each layer. At this frontend stage, the representation is device agnostic; however, we can provide useful statistics, such as the operation count and the memory footprint of the weights and activations of each layer.
In addition to loading topological information, \finnr may optionally load a set of trained weights. These weights are reordered and forwarded to subsequent \finnr stages for processing and inclusion in the final deployed network.

\subsection{The Intermediate Representation}
\label{sec:IR}
As is common practice \cite{sharma2016dnnweaver, venieris2016fpgaconvnet,
abadi2016tensorflow}, \finnr represents a QNN as a directed acyclic
graph. Its nodes represent layers and edges carry outputs from one
layer to become inputs to another. The key differentiator of \finnr's
intermediate representation (IR) is its quantization-awareness. Each
node is tagged with the quantization of its inputs, parameters
(weights) and outputs to enable quantization-aware optimizations (such as the
streamlining optimization described below) and the
mapping to backend primitives optimized for quantized computation.
Internally, the IR differentiates \textit{backend-neutral}
and \textit{backend-specific} layers.
When a QNN is initially imported into \finnr, its abstract
computational structure is solely represented by backend-neutral
layers. This representation is made backend-specific for a hardware
implementation by a series of transform and analysis passes. In the
derived graph, the network is decomposed into concrete hardware
building blocks, such as the SWU an the MVU.

\paragraph{\finnr Transform and Analysis Passes}
\label{secPasses}
\finnr employs \emph{passes}, i.e. small subprograms that
operate on the IR.
Each pass consumes an IR graph, and may (a) \emph{transform} the QNN
to output a modified graph, (b) \emph{analyze} the QNN to produce
metadata about its properties, or do both.
A pass may be composed of smaller passes to facilitate reuse and modularity.
We highlight some of the key passes implemented in \finnr below.

\paragraph{Direct Quantization}
The first and last layers of QNNs are often quantization-sensitive and
left using floating-point arithmetic \cite{cai:2017,
DBLP:journals/corr/ZhouNZWWZ16,
DBLP:journals/corr/RastegariORF16}. However, a modest direct
quantization as to 8-bit fixed-point quantities has little or no impact
on accuracy \cite{jouppi2017datacenter} but already leads to significant
resource savings. \finnr's direct quantization pass applies this
transformation to non-quantized layers converting its parameters to
fixed-point values of the specified bit precision.
For quantizations below 8\,bits, retraining is highly recommended but
is not part of this pass.

\paragraph{Streamlining}
\label{secStreamlining}
The original FINN paper described how batch normalization parameters can be absorbed into thresholds via simple mathematical manipulation.
This can be further generalized into a \textit{streamlining pass} that
absorbs floating-point scaling factors into multi-level thresholds as
explored by Umuroglu and Jahre \cite{umuroglu2017streamlined}.
This is done by collapsing scaling layers in front of a quantization
layer into a single linear transform that is then merged entirely into the
quantization by updating its thresholds.
The mathematically equivalent output QNN eliminates the storage and
compute overhead of the subsumed floating-point scaling factors.
Finally, as the maximum operator commutes with monotone functions such
as the quantization used in QNNs, maxpool layers may be moved behind a
quantization layer. This decreases the required precision of the
comparators in the maxpool layer, resulting in further resource savings.

\paragraph{FPGA Resource Analysis}
\label{secPassResAnl}
Chooses and scales hardware operators to optimize the
network performance within a given resource budget. The corresponding
algorithm is detailed in \rsec{secDataflowBalance}, and is integrated
into \finnr as an analysis pass.

\paragraph{FPGA Dataflow Architecture Generation}
\label{secPassDataflowGen}
Generates a synthesizable DF implementation from the IR
after the concurrency annotation by the resource analysis pass.
Each layer is converted to an equivalent FPGA layer representing a
building block of the QNN library parametrized to the determined
parallelism. Finally, the corresponding HLS code is generated.

\paragraph{FPGA Multilayer Offload Schedule Generation}
\label{secPassMultiOffGen}
Generates an execution schedule targeting a given MO implementation, which is the sequence of layers as specified in the IR graph.

\subsection{Backends}
Backends are responsible for consuming the IR graph and backend-specific information to create a deployment package, and/or providing performance/resource estimates for two previously introduced hardware architectures (DF and MO).
The deployment package consists of parameter data for the QNN model, and the backend-specific code that executes the model, consisting of both runtime environment as well as an executable hardware design for targeting dataflow and multilayer offload architectures, as well as a selection of predefined platforms.

\subsection{Controlling Performance and Resource Utilization}
\label{secDataflowBalance}
\finnr exploits the concurrency potential in a given QNN to generate a solution,
which is scaled to utilize the committed resources optimally by tuning the
previously introduced concurrency parameters $P$ (PE duplication), $Q$
(SIMD scaling), and $M$ (multi-vector parallelization). All of them
allow to accelerate the computation of the respective layer whose
throughput grows proportionally. For a feasible schedule, we choose $Q$ as a factor of $C$, $P$ as a factor of $C'$,
and $M$ as a factor of $N'$. Finally, $A$ represents the compute complexity of each layer and $M$ its cumulative parallelism.

\begin{algorithm}
  \KwIn  {A[0..L-1]}
  \KwData{M[0..L-1]}
  
  candidate := MO\tcc*{default to a multilayer offload}
  M[0..L-1] := 1   \tcc*{minimal dataflow compute}
  \tcc{Adopt dataflow with greater compute parallelism as long as feasible.}
  \While{feasible(M)}{
    candidate := \{ DF, M \}\;
    idx    := max\_index \{ A[.]/M[.] \}\;
    M[idx] := next greater factor of $C[i]\cdot C'[i]\cdot N'[i]$\;
  }
  return candidate\;
  \caption{Data Flow Balancing by \finnr}
  \label{algScale}
\end{algorithm}
The minimal DF implementation chooses a scaling of $1$ for all parameters
and all layers. Its feasibility within the committed resources as
determined by the cost functions of \rsec{secQNNimpl} decides whether
a retreat to an MO architecture is necessary or the DF performance
scaling can be pursued. The balanced scaling of the layers
in a DF pipeline is the key capability of \finnr. Having
determined the compute requirements of all the layers, it
systematically widens the most pressing bottlenecks as shown
by \ralg{algScale} until the resources are exhausted. The cumulative
scaling factor determined for each layer is used to tile its
computation into corresponding factors $Q$, $P$ and $M$.

\finnr estimates the performance of its generated implementation based
on the chosen parallelism and the reported initialization interval of
the building blocks. The layer compute time is evaluated as the
quotient of its compute requirements and the attained concurrency.
A throughput characterization is also performed for the other building
blocks (SWU, MP) to identify when they become the bottleneck within the pipeline.




\section{Evaluation}\label{secShowCases}
We present and evaluate \finnr on a range of platforms, with a selection of neural networks, for both dataflow and multilayer offload architectures, and compare it with state of the art.

\subsection{Experimental Setup}
A broad range of platforms have been targeted from embedded to datacenter scale. Their details are summarized in \rtab{tabPlatforms}.
PYNQ-Z1 is an open-source board and easily available, Ultra96 is a board based on the ultra96.org hardware specification using the newer Zynq Ultrascale+ architecture, and AWS\,F1 is a FaaS node in the public AWS cloud infrastructure.
\begin{table}
  \caption{Platform Summary}
  \label{tabPlatforms}
  \resizebox{0.8\linewidth}{!}{
  \begin{tabular}{llrrrr}\toprule
    \multicolumn{1}{c}{\textbf{\textbf{Platform}}}&
    \multicolumn{1}{c}{\textbf{Part\,\#}}&
    \multicolumn{1}{c}{\textbf{Node}}&
    \multicolumn{1}{c}{\textbf{DDR}}&
    \multicolumn{1}{c}{\textbf{kLUTs}}&
    \multicolumn{1}{c}{\textbf{BRAMs}}\\\midrule
    \textbf{AWS\,F1} & XCVU9P-FLGB2104-2-I & 16\,nm & 64\,GB & 1,180 & 4,320 \\
    \textbf{Ultra96} & XCZU3EG-SBVA484-1 & 16\,nm & 2\,GB & 71 & 432 \\
    \textbf{PYNQ-Z1~\cite{ds190}} & XC7Z020-1CLG400C & 28\,nm & 512\,MB &53.2 & 280 \\\bottomrule
  \end{tabular}
	}
  \vspace*{-3ex}
\end{table}
We considered four different types of neural network topologies to evaluate how well the performance prediction works for different computing patterns. The exact specifications can be found at GitHub \cite{finn-r-git} and descriptions are given in the following paragraphs.

\paragraph{MLP-4, CNV-6} These two topologies are identical to the ones used
in FINN~\cite{umuroglu2017finn}. One is a classifying multilayer perceptron
built from four large fully connected layers, and the other one implements a CIFAR-10, GTSRB or SVHN classifier
and is inspired by BinaryNet \cite{courbariaux:2016} and VGG-16 \cite{DBLP:journals/corr/SimonyanZ14a}.
\emph{Tincy\,YOLO} is a quantized adaptation of
Tiny\,YOLO, which itself is a stripped-down version of YOLO directly
provided by its authors~\cite{redmon:2016}. Tiny\,YOLO is described completely
in darknet~\cite{darknet}. Specifically, the differences
between Tincy\,YOLO and Tiny\,YOLO are as follows:
\begin {enumerate*} [1) ]%
\item the first and last convolution layers are converted to use 8-bit weights;
\item the other convolutional layers are converted to use 1-bit weights, 3-bit
activations;
\item the activation function is removed from the last convolutional layer;
\item all other activation functions converted from leaky-ReLU to ReLU;
\item increase the size of the number of OFMs in the 2nd convolutional layer
from 32 to 64;
\item a decrease in the number of OFMs of the 7th and 8th convolutional layers
from 1024 to 512;
\item removal of the first maxpool layer; and
\item an increase of stride in the first convolutional layer from 1 to 2.
\end {enumerate*}
The end result of these topological changes is a reduction in operations of
36\%, while reducing the mean average precision (mAP) accuracy from 57.1\% to
50.1\%. \emph{DoReFa-Net/PF}
The DoReFa-Net/PF topology is based off the DoReFa-Net topology described by
Zhou~et\,al.~\cite{DBLP:journals/corr/ZhouNZWWZ16}. DoReFa-Net itself, is a variant of
AlexNet~\cite{Krizhevsky:2012:ICD:2999134.2999257} with 1-bit weights, 2-bit
activations used in all layers except the first and last, where floating-point
weights are used. We further modify DoReFa-Net as follows:
\begin {enumerate*} [1) ]%
\item the first and last layer are converted from floating-point weights to
8-bit weights; and
\item the first 3 layers are pruned by 30\%, specifically, decreasing the
number of OFMs in these layers from 96, 128 and 384 to 68, 90 and 272
respectively.
\end {enumerate*}
The end result of the these topological changes is a reduction in operations of
56\%, while improving the overall accuracy (Top-1) by 0.2\%.
For completeness, the Top-5 accuracies we were able to achieve with DoReFa-Net/PF,
DoReFa-Net\,[\qnn{1}{2}] and DoReFa-Net\,[\qnn{FP}{FP}] was 74.0\%, 73.1\%
and 77.4\% respectively.

In terms of training, a combination of the following techniques were used to train
the networks described in \rtab{tabWorkloads}:
\begin {enumerate*} [1) ]%
\item the MLP-4 and CNV-6 networks were trained using the techniques
described by Courbariaux~et\,al.~\cite{courbariaux:2016};
\item DoReFa-Net/PF was trained using the techniques described by
Zhou~et\,al.~\cite{DBLP:journals/corr/ZhouNZWWZ16}, with the exception of the
\qnn{8}{8} layers, which are trained using the techniques described by
Su~et\,al.~\cite{su2018accuracy};
\item The YOLO networks were trained using the methods described by
Rastegari~et\,al.~\cite{DBLP:journals/corr/RastegariORF16} for the network weights,
Zhou~et\,al.~\cite{DBLP:journals/corr/ZhouNZWWZ16} for the activations and
Su~et\,al.~\cite{su2018accuracy} for the \qnn{8}{8} layers.
\end {enumerate*}
The pruning methods applied to realise Tincy\,YOLO and DoReFa-Net/PF are described
by Faraone~et\,al.~\cite{faraone2018hardware}.

The computational workloads of all these four QNN
variants are indicated by \rtab{tabWorkloads}. The stated figures
sum up all the additions and multiplications within the dot products
computed as an individual input frames passes through the
convolutional layers and perceptrons of a topology. The very small
MLP-4 network already asks for 6~million operations for each
image. Being used for a digit recognition, which is purely based on
shape, it is also the only network implementation that has been fully
binarized starting from the very input.

All the other topologies feature layers of a higher 8-bit precision,
particularly in order to interface naturally with their application
scenarios. This precision allows to easily convey the color
information of their visual inputs into their first layers. It is also
the natural choice for reporting confidence levels of complex
classification tasks. While the input and output layers, thus, pose
intrinsic quantization bounds, the hidden layers can often be
quantized significantly while practically maintaining accuracy. As
shown in \rtab{tabWorkloads}, the amount of computation that could be reduced
this way is enormous and accounts for the vast majority of the compute
in these networks.

For reference, we've also provided the accuracy and workloads of unpruned and/or
unquantized versions of these networks if they are available. No floating point
results are provided for the CNV-6 topology, as this network was designed specifically
as a BNN by Umuroglu~et\,al.~\cite{umuroglu2017finn}.
With regard to accuracy, we note that the difference in accuracy between MLP-6,
Tincy\,YOLO and DoReFa-Net/PF and their unquantized, unpruned equivalents is 1.05\%,
7.0\% and 5.6\% respectively. Whether these accuracies are still acceptable is dependent
on the target application.

\begin{table}
  \caption{Dot-Product Workloads \& Network Accuracies}
  \label{tabWorkloads}
  \resizebox{\linewidth}{!}{
  \begin{tabular}{lr@{\,}ccccccc}\toprule
		& \textbf{bin. Ops/Frame} & \textbf{High Precision Ops/Frame} & \textbf{Total Ops/Frame} & \textbf{Parameters} & \textbf{Accuracy} & \textbf{Dataset}\\\midrule
    \textbf{MLP-4}        &   6.0\,M [\qnn{1}{1}]&      --              &   6.0\,M & 2.9\,M & 97.69(Top-1 (\%)) & MNIST\\
    \textbf{MLP-4}        & --                   &6.0\,M [\qnn{FP}{FP}] &   6.0\,M & 2.9\,M & 98.74(Top-1 (\%)) & MNIST\\
    \textbf{CNV-6}        & 115.8\,M [\qnn{1}{1}]&  3.1\,M [\qnn{1}{8}] & 118.9\,M & 1.4\,M & 80.10(Top-1 (\%)) & CIFAR-10\\
    \textbf{CNV-6}        & 115.8\,M [\qnn{1}{1}]&  3.1\,M [\qnn{1}{8}] & 118.9\,M & 1.4\,M & 98.08(Top-1 (\%)) & GTSRB\\
    \textbf{CNV-6}        & 115.8\,M [\qnn{1}{1}]&  3.1\,M [\qnn{1}{8}] & 118.9\,M & 1.4\,M & 94.90(Top-1 (\%)) & SVHN\\
    \textbf{Tincy\,YOLO}   &4385.9\,M [\qnn{1}{3}]& 59.0\,M [\qnn{8}{8}] &4444.9\,M & 6.4\,M & 50.1(mAP (\%)) & VOC 2007\\
    \textbf{Tiny\,YOLO}    &6780.5\,M [\qnn{1}{3}]&194.9\,M [\qnn{8}{8}] &6975.3\,M & 15.9\,M& 48.7(mAP (\%)) & VOC 2007\\
    \textbf{Tiny\,YOLO}    & --                 &6975.3\,M [\qnn{FP}{FP}]&6975.3\,M & 15.9\,M& 57.1(mAP (\%)) & VOC 2007\\
    \textbf{DoReFa-Net/PF}&2009.7\,M [\qnn{1}{2}]&171.3\,M [\qnn{8}{8}] &2181.0\,M & 60.2\,M& 50.3(Top-1 (\%)) & ImageNet\\
    \textbf{DoReFa-Net}   &3626.3\,M [\qnn{1}{2}]&250.1\,M [\qnn{FP}{FP}] &3876.3\,M & 61.0\,M& 50.1(Top-1 (\%)) & ImageNet\\
    \textbf{DoReFa-Net}   & --                 &3876.3\,M [\qnn{FP}{FP}]&3876.3\,M & 61.0\,M& 55.9(Top-1 (\%)) & ImageNet\\\bottomrule
  \end{tabular}
  }
  \vspace*{-3ex}
\end{table}



\subsection{Measured Results and Evaluation}
Concrete implementations of the four described networks on the introduced platforms
were dimensioned, synthesized and measured. All results are summarized in \rtab{tabImplementations} together with prior art.
The datapoints targeting PYNQ-Z1 can be reproduced by using the open-source releases \cite{QNN-MO-PYNQ}, \cite{BNN-PYNQ}.
We constrain our comparison to only measured results
and exclude extrapolated numbers (\cite{nurvitadhi2017can},\cite{moss2017high},\cite{nurvitadhi:2016}\cite{DBLP:journals/corr/abs-1709-01134}).
For undisclosed CNN and MLP topologies in related work, we use the term CNV-X and MLP-X respectively.
Power and efficiency are measured in respect to board power.
HP and FP stand for 16-bit and 32-bit (half precision and full precision) floating point respectively.
Resources are only given for BRAM18s and LUTs on Xilinx devices where available.
URAM is excluded as many of the utilized devices do not support them, and optimal selection of LUTRAM, BRAM and URAM through the tool flow is not yet controlled.
DSP counts are not relevant for the extreme QNNs apart from the 8\,bit precision layers.
All implementations are compared in regards to performance, power, efficiency (=performance/power) for a specific network on a specific platform.
Where possible, we provide details on resource usage, and data types used.

\begin{table}
  \caption{\finnr Implementations and Prior Work}
  \label{tabImplementations}
\resizebox{\textwidth}{!}{
  \begin{tabular}{llrrrrrr@{\;}r}\toprule
    \textbf{CNN} & \textbf{Platform} 
    & \textbf{Clock} & \textbf{BRAM18} & \textbf{LUTs} & \textbf{Perf.(predicted)} & \textbf{(Power)} & \textbf{Efficiency} & \textbf{Precision} \\
    && (MHz)    &      &    & (GOp/s)(\%) & (W) & (GOp/s/W) & (\%) \\\midrule
\rowcolor[gray]{.9}\multicolumn{9}{l}{\textbf{\finnr Results}}\\
    \textbf{\finnr MLP-4} &\textbf{AWS\,F1} (DF) & 232.9   & 1,652 & 337,753 & 50,776 (79\%) & - & - & \qnn{1}{1}\\
    \textbf{\finnr MLP-4} &\textbf{Ultra96} (DF)& 300 & 417 &  38,205 & 5,110 (75\%) & 11.8 & 433 & \qnn{1}{1}\\
    \textbf{\finnr MLP-4} &\textbf{PYNQ-Z1} (DF)& 100 & 220 & 25,358 &  974 (99\%) & 2.5 & 390 & \qnn{1}{1} \\
    \textbf{\finnr CNV-6} &\textbf{AWS\,F1} (DF)& 237 &  1,888 &   332,637  &  12,109 (98\%) & - & - &\qnn{1}{1}\\
    \textbf{\finnr CNV-6}&\textbf{Ultra96} (DF)& 300 & 283 &   41,733 & 2,318 (99\%) & 10.7 & 217 & \qnn{1}{1}\\
    \textbf{\finnr CNV-6}&\textbf{PYNQ-Z1} (DF)& 100 & 242 &  25,770 &  341 (99\%) & 2.25 & 152 & \qnn{1}{1}\\
    \textbf{\finnr Tincy\,YOLO}&\textbf{AWS\,F1} (DF)& 249.8 & 2,638 & 242,457 & 5,271 (93\%) & - & - & \qnn{1}{3}\\
    \textbf{\finnr Tincy\,YOLO}&\textbf{Ultra96} (MO)& 220 & 316 & 40,808 & 288 (68\%) & 9.7 & 30 & \qnn{1}{3}\\
    \textbf{\finnr Tincy\,YOLO}&\textbf{PYNQ-Z1} (MO)& 100 & 280 & 46,507 &  133 (66\%) & 2.5 & 53 & \qnn{1}{3} \\
    \textbf{\finnr DoReFa-Net/PF}&\textbf{AWS\,F1} (DF)& 155.8&  1,332  &  476,970 & 11,431 (92\%) & - & - & \qnn{1}{2}\\
    \textbf{\finnr DoReFa-Net/PF}&\textbf{Ultra96} (MO)& 220 & 432 &36,249&  400 (70\%) & 10.2 & 39 & \qnn{1}{2}\\
    \textbf{\finnr DoReFa-Net/PF}&\textbf{PYNQ-Z1} (MO)& 100 & 278 & 35,657 &  129 (48\%) & 2.5 & 51 & \qnn{1}{2}\\
\rowcolor[gray]{.9}\multicolumn{9}{l}{\textbf{Related Work - Reduced Precision}}\\
\textbf{FINN MLP-4~\cite{umuroglu2017finn}} & \textbf{ZC706} & 200 & 396 & 82,988 & 9,086(-) & 22.6 & 402 & \qnn{1}{1}\\
\textbf{FINN CNV-6~\cite{umuroglu2017finn}} & \textbf{ZC706} & 200 & 186 & 46,253 & 2,466(-) & 11.7 & 210 & \qnn{1}{1} \\
\textbf{FINN CNV-6} & \textbf{ADM-PCIE-8K5} & 125 & 1,659 & 365,963 & 14,814(-) & $\sim41$ & 361 & \qnn{1}{1} \\ 
\textbf{Alemdar~et\,al.~\cite{alemdar2016ternary} CNV-X} & \textbf{Kintex-7 160T}  & - & - & - & $\sim878$(-) & - & - & \qnn{2}{2} \\ 
\textbf{Prost-Boucle~et\,al.~\cite{prost2017scalable} CNV-6} & \textbf{VC709} & & & & $\sim3,029$(-) &
6.64$^*$ 
& - & \qnn{2}{2}\\
\textbf{Park and Sung~\cite{park2016fpga} MLP-X} & \textbf{ZC706}  & 172 & - & - & $\sim210$(-) & $\sim5$ & 42 & \qnn{3}{8} \\
\textbf{Nakahara~et\,al.~\cite{nakahara2017fully} CNV-X} & \textbf{Zedboard} & 144 & 32$^\dagger$ & 18,325$^\dagger$ & 329(-) & 2.3 & 143 & \qnn{1}{1} \\
\textbf{Yonekawa~et\,al.~\cite{yonekawa2017chip} CNV-X} & \textbf{ZCU102} & 150 & 1,367$^\dagger$ & $\sim22,095$$^\dagger$ & 461(-) & 22 & 21 & \qnn{1}{1} \\
\textbf{Zhao~et\,al.~\cite{zhao2017accelerating} CNV-X} & \textbf{Zedboard} & 143 & 94$^\dagger$ & $\sim46,900$$^\dagger$ & 207.8(-) & 4.7 & 44 & \qnn{1}{1} \\
\textbf{Jiao~et\,al.~\cite{jiao2017accelerating} DoReFa-Net} & \textbf{Zynq 7020} & 200 & 106$^\dagger$ & $\sim44,000$$^\dagger$ & 410.2(-) & 2.3 & 182 & \qnn{1}{2} \\
\textbf{Liang~et\,al.~\cite{liang2017fp} CNV-X} & \textbf{MPC-X2000 DFE} & 150 & - & - & 9,396(-) & 26.2 & 359 & \qnn{1-8}{1-8}\\
\rowcolor[gray]{.9}\multicolumn{3}{l}{\textbf{Related Work - Higher Precision}}
&\multicolumn{6}{r}{\footnotesize $^*$ no full board power, only as measured by PMBUS available rails; $^\dagger$ includes shell overhead}\\
\textbf{Wei~et\,al.~\cite{wei2017automated} AlexNet} &\textbf{AWS\,F1} & 230 & 1,682 & 545,000 & 1,884.2(-) & - & - & \qnn{8}{16} \\
\textbf{Wei~et\,al.~\cite{wei2017automated} VGG16} &\textbf{AWS\,F1}   & 240 & 1,690 & 504,000 & 2,260.9(-) & - & - &\qnn{8}{16} \\
\textbf{Wei~et\,al.~\cite{wei2017automated} ResNet} &\textbf{AWS\,F1}  & 230 & 1,644 & 496,000 & 1,875.3(-) & - & - & \qnn{8}{16} \\
\textbf{Zhang~et\,al.~\cite{zhang2016caffeine} VGG16} &\textbf{VC709} & 150 & - & - & 354(-) & 26 & 14 & \qnn{16}{16}  \\
\textbf{Aydonat~et\,al.~\cite{aydonat2017opencl} AlexNet} & \textbf{GX1150} & 303 & - & - & 1,382(-) & $\sim45$ & 31 & \qnn{HP}{HP} \\
\textbf{Zhang~et\,al.~\cite{zhang2017improving} VGG16} & \textbf{GX1150}    & 385 & - & - & 1,790(-) & $\sim37.46$ & 48 &\qnn{16}{16} \\
\textbf{Hedge~et\,al.~\cite{caffepresso} Caltech101} & \textbf{ZC706} & 180 & - & - & 11.5(-) & 19 & $<1$ & \qnn{16}{16} \\
\textbf{Ma~et\,al.~\cite{ma2017optimizing} VGG16} & \textbf{GX1150}   & 150 & - & - & 645.25(-) & - & - & \qnn{8}{16} \\
\textbf{Ma~et\,al.~\cite{ma2017automatic} NiN} & \textbf{GX1150} & 200 & - & - & 588(-) & - & - & \qnn{16}{16} \\
\textbf{Ma~et\,al.~\cite{ma2017automatic} VGG-16} & \textbf{GX1150} & 200 & - & - & 720(-) & - & - & \qnn{16}{16} \\
\textbf{Ma~et\,al.~\cite{ma2017automatic} ResNet-50} & \textbf{GX1150} & 200 & - & - & 619(-) & - & - & \qnn{16}{16} \\
\textbf{Ma~et\,al.~\cite{ma2017automatic} ResNet-152} & \textbf{GX1150} & 200 & - & - & 710(-) & - & - & \qnn{16}{16} \\
\bottomrule \\
\multicolumn{9}{l}{Metrics not reported by prior work are indicated by -. Estimated values are denoted as \textasciitilde.} \\

\end{tabular}
}
  \vspace*{-3ex}
\end{table}

There are a number of key observations that can be taken from the table:
First of all, as expected there is a direct correlation between precision and achievable peak performance and energy efficiency, clearly demonstrating the value of QNNs.
Wei~et\,al.~\cite{wei2017automated} as well as recent Xilinx demonstration at SC'2017, clearly show close to possible peak performance for higher precision implementations on AWS\,F1.
These solutions are bottlenecked by DSP availability. We show here that QNNs can provide further performance scalability by leveraging the vast amount of LUTs.
This can be easily seen when comparing peak performance and efficiency on the same platform as for example AWS\,F1 across a spectrum of precisions.
Secondly, CNNs compared to MLPs achieve on the same platform lower peak performance. This is because the actual compute in form of multiply accumulate has a lower percentage. As shown with the cost functions, CNNs contain also SWUs and  MPs and the overhead has clearly an impact on overall performance.
Compared to the original FINN, we demonstrate that the performance can scale with the size of the FPGA. Many new network topologies and platforms and different precision implementations were added.

The best performance numbers are achieved for a fully binarized MLP on
AWS\,F1, at currently 50\,TOp/s and 12\,TOp/s for a CNN. Note
that these are initial results with much yet to gain
by investing into timing closure for instance.
The best efficiency can only be reported for the embedded platforms as for other platforms board-level power measurements are not available.
The highest value was achieved for \finnr on the MLP on the Ultra96 platform with 433\,GOp/s/W.

While DorefaNet is no longer state-of-the art in regards to accuracy, the network is still representative and its implementation demonstrates that image classification with large networks on large input sizes can be highly performant on FPGA-based implementations leveraging reduced precision. The implementation defaulted to 155.8\,MHz and achieved with that a measured 11.43\,TOp/s performance. Timing closure at 300\,MHz would immediately translate into a performance of 22\,TOp/s as the implementation is not memory bound. Furthermore, there is still scope to scale spatially.

Compared to other reduced precision implementations on Virtex and Kintex platforms,
we are higher performing on the same network compared to Prost~et\,al.~\cite{prost2017scalable} on larger platforms (efficiency cannot be compared as not the full board power is taken into consideration).
We outperform Alemdar~et\,al.~\cite{alemdar2016ternary} by over an order of magnitude.
Another highlight amongst related work is shown by the FP-BNN project from Liang~et\,al.~\cite{liang2017fp} which reports up to 9.4\,TOp/s overall performance on a server class platform across a selection of networks with mixed precision.
In regards to ZC706, our solution provides a speed-up of 10$\times$ and higher on the basis of dataflow-based architectures.
ZedBoard data points are best compared to Pynq-Z1, as they feature the same device.
In comparison to our dataflow architecture, Nakahara~et\,al.~\cite{nakahara2017fully} and Zhao~et\,al.~\cite{zhao2017accelerating}
are up to 2$\times$ slower. The binary network accelerator by Yonekawa~et\,al.~\cite{yonekawa2017chip} is hard to compare as it was done on a ZCU102, where the FPGA device is around 4$\times$ larger than the same device on the Ultra96. Even with the significantly smaller device, the CNV based \finnr implementation is 4$\times$ faster.
Finally, Jiao~et\,al.~\cite{jiao2017accelerating} report excellent performance for a large network on a small device with 410.2\,GOp/s.
Due to the large size of this network, \finnr cannot use a DF design, and MO achieves 129\,GOp/s.
Overall, we believe that \finnr delivers highly competitive performance and efficiency, with room for further improvements.

In regards to the framework, the broad spectrum of networks, platforms and implementations demonstrates the flexibility that the tool can provide through automated scaling.
Furthermore, the performance estimations for DF are 75\% - 99\% accurate and as such a good indication of achievable performance without having to go through a full implementation flow. Also note that the prediction is around the 50\% mark of the original AWS\,F1 roofline, and as such more accurate.
The performance estimations for the MO still differ from actual implementations. This is due to the more complex nature of the system, which involves
significantly more external memory transfers for pulling weights and FM, which latencies are currently estimated using a peak-bandwidth model.
In the case of DoReFa-Net with MO on PYNQ-Z1, the measured performances are largerly lower than expected (48 \%).
This is due to memory re-organization performed in the host whenever a split layer is present to concatenate feature maps for the following layer. The execution time scales with number of available threads on the host.
This non-ideal behaviour is overcome in the Ultra96 case, which features 4 cores against the 2 available on PYNQ-Z1.

\section{Conclusions \& Outlook}\label{secCon}
Quantization and reduced precision representations for inputs, activations and weights provide a promising approach to
scalability in performance, power efficiency and storage footprint for CNNs. They create interesting design
trade-offs in exchange for a small reduction in accuracy. Especially in conjunction with FPGAs,
a broad spectrum of architectures for inference engines can be explored whereby the minimal precision to achieve precisely the
numerical accuracy required for a particular application can be exploited.

In this article, we described \finnr, an end-to-end framework that automates the exploration of customized hardware accelerators.
\finnr is the second generation of the FINN tool, extending the original with support for arbitrary precision
and more flexibility in the end architecture and target platforms, including hardware cost estimatation for given devices.
We evaluated the generated architectures on four different reduced precision neural networks,
from small CIFAR-10 classifiers to a YOLO-based object detection on PASCAL VOC datasets
on a range of platforms including PYNQ and AWS\,F1. The generated design end-points showcase the promise of QNNs, demonstrating 5\,TOp/s on
embedded platforms as well as 50\,TOp/s on datacenter platforms.
The broad selection of measured results validates the workflow and the  flexibility of the framework.

The spectrum of possible design space tradeoffs is vast, and while we hope to shed some light on what is possible, many variants remain unexplored.
We are currently in the process of open-sourcing the framework, such that the broader community can help investigate the design space.
The networks themselves as well as a subset of implementations can already be found on github \cite{finn-r-git}.
From a research point of view, we are focusing on tightening up the performance and resource predictions,
adding support for more networks such as the promising residual networks and LSTM, and providing more automation within the framework itself, while a separate strand of research is working on improving accuracy through novel training techniques and quantization methods with different numerical representations.

\begin{acks}
  \begin{tabularx}{\linewidth}{@{}cX@{}}
  \raisebox{-7.8mm}{\resizebox{!}{1cm}{
\begin{tikzpicture}
\fill[fill={rgb,255:red,0;green,51;blue,153}] (-27,-18) rectangle (27,18);
\pgfmathsetmacro\inr{tan(36)/cos(18)}
\foreach \i in {0,1,...,11} {
  \begin{scope}[shift={(30*\i:12)}]
    \fill[fill={rgb,255:red,255;green,204;blue,0}] (90:2)
	\foreach \x in {0,1,...,4} { -- (90+72*\x:2) -- (126+72*\x:\inr) };
  \end{scope}
}
\end{tikzpicture}
}}&\small
  This project has received funding from the \grantsponsor{EU}{European Union}{}'s Framework
  Programme for Research and Innovation Horizon 2020 (2014-2020) under the
  \grantnum{EU}{Marie Sk{\l}odowska-Curie Grant Agreement No.\,751339}.
  Miriam Leeser is supported in part by the National Science Foundation under Grant No. 1717213.
  \end{tabularx}
\end{acks}

\bibliographystyle{ACM-Reference-Format}
\bibliography{literature}

\end{document}